\documentclass[aps,pra,reprint,superscriptaddress]{revtex4-2}
\usepackage{amsmath}
\usepackage{amssymb}
\usepackage{braket}
\usepackage{physics}
\usepackage{mathtools}

\newcommand{\ie}{\textit{i}.\textit{e}.}
\newcommand{\eg}{\textit{e}.\textit{g}.}
\newcommand{\apriori}{\textit{a priori}}
\newcommand{\rhohat}{\hat{\rho}}
\newcommand{\Ehat}{\hat{E}}
\newcommand{\Ehatyth}{\Ehat_{y|\btheta}}

\newcommand{\qhat}{\hat{q}}
\newcommand{\qhati}{\qhat_i}
\newcommand{\btheta}{\boldsymbol{\theta}}
\newcommand{\yx}{y_x}
\newcommand{\yp}{y_p}
\begin{document}

\title{Estimation of Gaussian random displacement using non-Gaussian states}

\author{Fumiya Hanamura}
\affiliation{Department of Applied Physics, School of Engineering, The University of Tokyo, 7-3-1 Hongo, Bunkyo-ku, Tokyo 113-8656, Japan}
\author{Warit Asavanant}
\affiliation{Department of Applied Physics, School of Engineering, The University of Tokyo, 7-3-1 Hongo, Bunkyo-ku, Tokyo 113-8656, Japan}
\author{Kosuke Fukui}
\affiliation{Department of Applied Physics, School of Engineering, The University of Tokyo, 7-3-1 Hongo, Bunkyo-ku, Tokyo 113-8656, Japan}
\author{Shunya Konno}
\affiliation{Department of Applied Physics, School of Engineering, The University of Tokyo, 7-3-1 Hongo, Bunkyo-ku, Tokyo 113-8656, Japan}
\author{Akira Furusawa}
\email{akiraf@ap.t.u-tokyo.ac.jp}
\affiliation{Department of Applied Physics, School of Engineering, The University of Tokyo, 7-3-1 Hongo, Bunkyo-ku, Tokyo 113-8656, Japan}
\affiliation{Optical Quantum Computing Research Team, RIKEN Center for Quantum Computing, 
2-1 Hirosawa, Wako, Saitama, 351-0198, Japan}

\date{\today}

\begin{abstract}
In continuous-variable quantum information processing, quantum error correction of Gaussian errors requires simultaneous estimation of both quadrature components of displacements on phase space. However, quadrature operators $x$ and $p$ are non-commutative conjugate observables, whose simultaneous measurement is prohibited by the uncertainty principle. Gottesman-Kitaev-Preskill (GKP) error correction deals with this problem using complex non-Gaussian states called GKP states. On the other hand, simultaneous estimation of displacement using experimentally feasible non-Gaussian states has not been well studied. In this paper, we consider a multi-parameter estimation problem of displacements assuming an isotropic Gaussian prior distribution and allowing post-selection of measurement outcomes. We derive a lower bound for the estimation error when only Gaussian operations are used, and show that even simple non-Gaussian states such as single-photon states can beat this bound. Based on Ghosh's bound, we also obtain a lower bound for the estimation error when the maximum photon number of the input state is given. Our results reveal the role of non-Gaussianity in the estimation of displacements, and pave the way toward the error correction of Gaussian errors using experimentally feasible non-Gaussian states.
\end{abstract}


\maketitle

\section{Introduction}

Continuous-variable optical quantum information processing has attracted much attention for its extent scalability achieved by large-scale cluster states \cite{doi:10.1126/science.aay2645,doi:10.1126/science.aay4354} which enable universal Gaussian operations \cite{PhysRevApplied.16.034005,larsen2021deterministic}. As a next step for the realization of practical quantum information processing, quantum error correction \cite{gottesman2010introduction} is an essential key. Gaussian error is the most common type of error in optical systems. It includes photon losses and Gaussian quantum channel \cite{harrington2001achievable} which is defined as phase-space displacements following an isotropic Gaussian distribution. It has been proven, however, that Gaussian errors imposed on Gaussian states cannot be corrected using only Gaussian operations \cite{eisert2002distilling,niset2009no}. Therefore, non-Gaussian states play an crucial role in the optical quantum information processing.

Gottesman-Kitaev-Preskill (GKP) error correction \cite{gottesman2001encoding} is one of the promising ways to correct Gaussian errors using a single-mode states. It uses highly complex non-Gaussian states called GKP states to correct errors. The information regarding displacements on both $x$ and $p$ quadratures imposed on GKP states is extracted by homodyne measurements and ancillary GKP states in the error syndrome measurement \cite{gottesman2001encoding}. Although $x$ and $p$ are non-commutative conjugate observables whose simultaneous measurement is prohibited by the uncertainty principle, the non-Gaussianity of GKP states and the prior information that the displacement is small enable accurate estimation of both parameters. However, GKP states are difficult to experimentally generate due to their high non-Gaussianity. Although their generation has been reported in other physical systems \cite{fluhmann2019encoding,campagne2020quantum}, optical generation of GKP states have not been achieved yet. Several other protocols for correcting Gaussian errors using non-Gaussian states (called bosonic codes) are known \cite{Terhal_2020}, but it has not been unraveled how experimentally feasible and simpler non-Gaussian states such as Fock states can be exploited for quantum error correction.


In this paper, we investigate a multi-parameter quantum estimation \cite{personick1971application,yuen1973multiple,holevo2011probabilistic,paris2009quantum,helstrom1976quantum,braunstein1994statistical} problem of displacements using non-Gaussian states. We assume an isotropic Gaussian prior distribution of the parameters and post-select measurement outcomes. First we derive a lower bound for the estimation error when only Gaussian states and Gaussian operations are used. Then we show that this bound can be beaten for some range of the prior variance even with only simple non-Gaussian states such as single-photon states. This result reveals the role of non-Gaussianity in the estimation of displacements and opens up the possibility of correcting Gaussian errors using experimentally feasible non-Gaussian states. We also derive a lower bound of the estimation error depending on the maximum photon number of the input state based on Ghosh's bound \cite{ghosh1993cramer}.

We evaluate the estimation error using the mean square error averaged with respect to the posterior distribution with the measurement outcome fixed. This is because it corresponds to the natural situation where one knows the prior distribution of the parameters and can post-select ``good'' measurement outcomes that reduces the estimation error. Although this criteria is different from frequentist estimation error (which fixes the true value of the parameter and takes average with respect to the measurement outcomes) or the Bayesian estimation error (which takes average with respect to both the true value and the measurement outcomes), it is a commonly used method in Bayesian statistics \cite{e20090628}. Reference \cite{duivenvoorden2017single} also studies the simultaneous displacement estimation problem using non-Gaussian states, although they adopt frequentist estimation error. For Gaussian states, estimation of displacement is a classical problem and has many previous studies \cite{assad2020accessible,bradshaw2018ultimate,genoni2013optimal}. For single-parameter estimation using non-Gaussian states, there have been some recent researches  \cite{oh2020optical,mccormick2019quantum}.

In Sec.~\ref{section:estimation_problem}, we formulate the Gaussian displacement estimation problem. In Sec.~\ref{section:gaussian_bound}, we derive lower bounds for the estimation error when only Gaussian (classical) states and Gaussian operations are used. In Sec.~\ref{section:non_gaussian}, we show that the bounds derived in Sec.~\ref{section:gaussian_bound} can be beaten using non-Gaussian states. In Sec.~\ref{section:lower_bound}, a statistical lower bound on the estimation error which depends on the maximum photon number and the post-selection probability is obtained, based on Ghosh's bound. Sec.~\ref{sec:conclusion} summarizes our paper and introduces some future works.  

\section{Gaussian displacement estimation problem}\label{section:estimation_problem}
\begin{figure}[ht]
\centering
\includegraphics{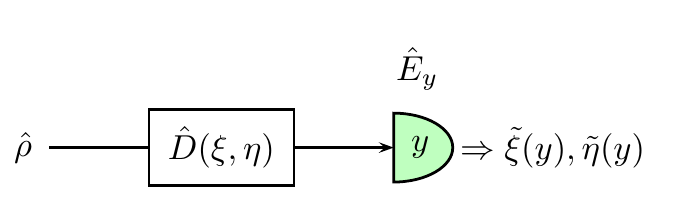}
\caption{A schematic representation of Gaussian displacement estimation problem. One estimates the amount of displacement $\xi,\eta$ using the measurement outcome $y$, which correspnds to a POVM element $\Ehat_y$. One can use prior information that $\xi,\eta$ follow an isotropic Gaussian distribution Eq.~(\ref{eq:gaussian_prior_dist}).}
\label{fig:displacement_estimation}
\end{figure}
Figure \ref{fig:displacement_estimation} depicts the Gaussian displacement estimation problem. We consider a single-mode bosonic state $\rhohat$ represented by quadrature operators $\hat{x},\hat{p}$ satisfying $\qty[\hat{x},\hat{p}]=i$. The phase space displacement operator $\hat{D}(\xi,\eta)$ is defined as
\begin{equation}
    \hat{D}(\xi,\eta)=\exp(i\eta\hat{x}-i\xi\hat{p}),
\end{equation}
and represents a displacement $(x,p)\to(x+\xi,p+\eta)$. First $\hat{D}(\xi,\eta)$ acts on $\rhohat$, then a measurement is performed, and the measurement outcome $y$ is obtained. This measurement corresponds to some POVM  (positive operator valued measure) $\{\Ehat_y\}$ satisfying $\int \Ehat_y dy=\hat{I}$, where $\hat{I}$ is an identity operator. We assume that $\xi,\eta$ are random variables following an isotropic Gaussian distribution with known variance and mean $0$:
\begin{equation}
    p(\xi,\eta)= \frac{1}{\pi v}\exp(-\frac{\xi^2+\eta^2}{v}).\label{eq:gaussian_prior_dist}
\end{equation}
Note that the mean square distance with respect to this distribution is $\expval{\xi^2}+\expval{\eta^2}=v$. Displacement following an isotropic Gaussian distribution corresponds to a common type of noise in bosonic systems called Gaussian quantum channel \cite{harrington2001achievable}, or additive Gaussian noise. The problem is to estimate $(\xi,\eta)$ from the value of $y$. We assume that one performs a Bayesian estimation using the \apriori{} information Eq.~(\ref{eq:gaussian_prior_dist}). The conditional probability density of obtaining $y$, when the values of $(\xi,\eta)$ are fixed, is given by
\begin{equation}
    p(y|\xi,\eta)=\Tr{\hat{D}(\xi,\eta)\rhohat\hat{D}^\dagger(\xi,\eta)\Ehat_y},\label{eq:cond_prob}
\end{equation}
where $\Tr{\cdot}$ denotes the trace operation. Defining the Wigner function $W_{\hat{A}}$ of an operator $\hat{A}$ by
\begin{equation}
    W_{\hat{A}}(x,p):=\frac{1}{2\pi}\int \exp(ipx')\Braket{x-\tfrac{x'}{2}|\hat{A}|x+\tfrac{x'}{2}} dx',
\end{equation}
Equation (\ref{eq:cond_prob}) can be expressed in terms of Wigner functions of $\rhohat$ and $\Ehat_y$ \cite{leonhardt2010essential}:
\begin{equation}
    p(y|\xi,\eta)=2\pi\int W_{\rhohat}(x-\xi,p-\eta)W_{\Ehat_y}(x,p) dxdp.\label{eq:cond_prob_wigner}
\end{equation}
Once the value of $y$ is known, one gets the corresponding posterior distribution of $\xi,\eta$:
\begin{equation}
    p(\xi,\eta|y)=\frac{p(y|\xi,\eta)p(\xi,\eta)}{p(y)},\label{eq:post_prob}
\end{equation}
where $p(y)=\int\int p(y|\xi,\eta)p(\xi,\eta) d\xi d\eta$ is the marginal probability distribution of $y$.
Suppose one estimates the values of $\xi,\eta$ as $\tilde{\xi}(y),\tilde{\eta}(y)$, corresponding to the value of $y$. To evaluate the accuracy of the estimation, the mean square error with respect to the joint probability distribution is one of the standard choices in Bayesian quantum estimation:
\begin{equation}
    v'_{\mathrm{Bayes}}:=\int p(\xi,\eta,y) \qty{(\xi-\tilde{\xi}(y))^2+(\eta-\tilde{\eta}(y))^2} d\xi d\eta dy,\label{eq:vp_bayes_definition}
\end{equation}
where $p(\xi,\eta,y)=p(y|\xi,\eta)p(\xi,\eta)$ is the joint probability distribution of $(\xi,\eta,y)$.

However, here we consider the post-selection of a specific measurement outcome $y$, and take average only with respect to the true value $\xi,\eta$, obtaining

\begin{equation}
    v':=\int p(\xi,\eta|y) \qty{(\xi-\tilde{\xi}(y))^2+(\eta-\tilde{\eta}(y))^2} d\xi d\eta.\label{eq:vp_definition}
\end{equation}
The meaning of this quantity can be considered as the expected amount of error after obtaining the outcome $y$. Note that more generally there are other options for the estimation error measure (\eg{} mean square error with a general weight matrix \cite{gill1995applications,tsang2020physics}), and here we choose this $v'$ for simplicity.
We consider the case $\tilde{\xi}(y),\tilde{\eta}(y)$ are chosen as the averages with respect to the posterior distribution:
\begin{align}
    \tilde{\xi}(y)&=\int p(\xi,\eta|y) \xi d\xi d\eta,\\
    \tilde{\eta}(y)&=\int p(\xi,\eta|y) \eta d\xi d\eta,
\end{align}
which is the optimal choice to minimize $v'$. For any probability density function $g$ satisfying $\int g(\xi,\eta)d\xi d\eta=1$ and $ g(\xi,\eta)>0$, we define $\Sigma[g(\xi,\eta)]$ as the covariance matrix of $(\xi,\eta)$ with respect to $g(\xi,\eta)$. Then, $v'$ can also be expressed as
\begin{equation}
    v'=\Tr{\Sigma[p(\xi,\eta|y)]}.\label{eq:vp_trace}
\end{equation}

Note that $v'$ depends on the measurement outcome $y$. Usually, in Bayesian quantum estimation, one evaluates the error of estimation by averaging over the measurement outcomes \cite{paris2009quantum}. However, in this paper, we allow post-selection of $y$ to get smaller $v'$. In Sec.~\ref{section:gaussian_bound}, we see that one can still obtain lower bounds on $v'$ with this assumption.

We also assume that there exists no initial entanglement between the input state and the measurement system.
The case where entanglements exist is studied in \eg{} Ref.~\cite{genoni2013optimal}, and it is shown that the amount of displacement can be estimated in arbitrary high precision if one uses two-mode squeezed vacuum with high squeezing level.
\section{Classical and Gaussian bounds}\label{section:gaussian_bound}
In this section, we derive lower bounds on the estimation error $v'$ (Eq.~(\ref{eq:vp_definition})), when only Gaussian states and Gaussian operations are allowed. We only have to consider the case where $\rhohat$ is a pure Gaussian state and $\Ehat_y\propto\ketbra{\psi}{\psi}$ for some pure Gaussian state $\ket{\psi}$. Indeed, suppose we have $\rhohat=t\rhohat_1+(1-t)\rhohat_2, 0<t<1$. Then $v'(\rhohat)\geq \min\qty{v'(\rhohat_1),v'(\rhohat_2)}$ because the variance is concave with respect to the probability distribution. In the same way, one can also show that mixing $\Ehat_y$ does not reduce $v'$ either. Wigner functions of pure Gaussian states can be written as
\begin{equation}
    W(x,p)=\frac{1}{\pi}\exp(-\frac{1}{2}(\boldsymbol{q}-\boldsymbol{\mu})^T\boldsymbol{\Sigma}(\boldsymbol{q}-\boldsymbol{\mu})),
\end{equation}
where $\boldsymbol{q}:=(x,p)^T$, and $\boldsymbol{\mu}$  and $\boldsymbol{\Sigma}$ are parameters representing the mean and the covariance matrix of the Gaussian function, and $\boldsymbol{\Sigma}$ satisfies $\det\boldsymbol{\Sigma}=\frac14$. Note that this constraint means that the Heisenberg's uncertainity is saturated by pure Gaussian states \cite{jackiw1968minimum}. Thus, Wigner functions of $\rhohat$ and $\Ehat_y$ are both Gaussian functions, and satisfy
\begin{equation}
    \det\Sigma[W_{\rhohat}(\xi,\eta)]=\det\Sigma[W_{\Ehat_y}(\xi,\eta)]=\frac{1}{4}\label{eq:uncertanity}.
\end{equation}
Here we extend our definition of $\Sigma$ in the last section to unnormalized distribution functions by defining $\Sigma[g(\xi,\eta)]:=\Sigma[\bar{g}(\xi,\eta)]$ for any unnormalized distribution function $g(\xi,\eta)$ satisfying $\int g(\xi,\eta)d\xi d\eta<\infty$ and $ g(\xi,\eta)>0$, where $\bar{g}(\xi,\eta)=Ng(\xi,\eta)$ is the normalization of $g$ which satisfies $\int \bar{g}(\xi,\eta)d\xi d\eta=1$. Note that although generally Wigner functions can have negative parts, Gaussian Wigner functions are positive and thus can be regarded as probability distributions. Using Eq.~(\ref{eq:cond_prob_wigner}) and a property of convolution, $p(y|\xi,\eta)$ is also Gaussian with respect to $(\xi,\eta)$, and its covariance matrix is
\begin{equation}
    \Sigma[p(y|\xi,\eta)]=\Sigma[W_{\rhohat}(\xi,\eta)]+\Sigma[W_{\Ehat_y}(\xi,\eta)].\label{eq:variance_additivity}
\end{equation}
Furthermore, from Eq.~(\ref{eq:post_prob}), $p(\xi,\eta|y)$ is also Gaussian, and
\begin{equation}
    \Sigma[p(\xi,\eta|y)]^{-1}=\Sigma[p(y|\xi,\eta)]^{-1}+\Sigma[p(\xi,\eta)]^{-1}\label{eq:variance_harmonic_mean}
\end{equation}
holds.
\subsection{Classical bound}
First we consider the case when only classical states, \ie{} coherent states, are available as $\rhohat$. In this case, we have $\Sigma[W_{\rhohat}(\xi,\eta)]=\frac{1}{2}I$, where $I$ is the identity matrix. From Eq.~(\ref{eq:uncertanity}), two eigenvalues of $\Sigma[W_{\Ehat_y}(\xi,\eta)]$ can be written as $a/2,1/(2a)$ for $a>0$. Then, from Eqs.~(\ref{eq:vp_trace}), (\ref{eq:variance_additivity}) and (\ref{eq:variance_harmonic_mean}), one obtains
\begin{equation}
    v'=\frac{1}{2/v+2/(1+a)}+\frac{1}{2/v+2a/(1+a)}.
\end{equation}
This is minimized when $a=1$, regardless of the value of $v$. Therefore, in this case, optimal $\Ehat_y$ is a projection to a classical state, and the classical lower bound of $v'$ is given by
\begin{equation}
    v'_{\mathrm{C}}=\frac{2v}{v+2}.\label{eq:classical_bound}
\end{equation}

\subsection{Gaussian bound}
Next we consider the case where arbitrary Gaussian states and operations can be used. For positive definite $2\times2$ matrices $A,B$,
\begin{equation}
    \det(A+B)\geq\det A+\det B +2\sqrt{\det A\det B}\label{eq:det_ineq}
\end{equation}
holds (See Appendix \ref{section:det_ineq} for the derivation). Using this fact and Eqs.~(\ref{eq:uncertanity}) and (\ref{eq:variance_additivity}), one can show that
\begin{equation}
    \det\Sigma[p(y|\xi,\eta)]\geq 1.\label{eq:gaussian_condition}
\end{equation}
Conversely, any Gaussian $p(y|\xi,\eta)$ satisfying Eq.~(\ref{eq:gaussian_condition}) is possible by taking $\Sigma[W_{\rhohat}(\xi,\eta)]=\Sigma[W_{\Ehat_y}(\xi,\eta)]=\Sigma[p(y|\xi,\eta)]/2$. Because $p(y|\xi,\eta)$ minimizing $v'$ must saturate Eq.~(\ref{eq:gaussian_condition}), its two eigenvalues can be written as $a,1/a$ for some $a>0$, and
\begin{equation}
    v'=\frac{1}{2/v+a}+\frac{1}{2/v+1/a}
\end{equation}
holds. The value of $a$ which minimizes $v'$ depends on the value of $v$. When $v>2$, $a=1$ minimizes $v'$. This means that choosing vacuum (or coherent) states for both of $\rhohat$ and $\Ehat_y$ is optimal, hence the minimum of $v'$ coincides with the classical bound, $\frac{2v}{v+2}$. On the other hand, when $v<2$, by taking $a\to0$ or $a\to\infty$, \ie{} taking both $\rhohat$ and $\Ehat_y$ to be infinitely squeezed states, $v'$ approaches to the lower bound $v/2$. It is precisely half of the original variance $v$, because variance along the squeezed direction is made $0$, and the variance along the antisqueezed direction remains $v/2$. To summarize, the Gaussian lower bound of $v'$ is given by
\begin{equation}
    v'_{\mathrm{G}}=\begin{dcases}\frac{v}{2}&(v<2)\\\frac{2v}{v+2}&(v>2)\end{dcases}.\label{eq:gaussian_bound}
\end{equation}

Note that the lower bounds of the Bayesian estimation error $v'_{\mathrm{Bayes}}$ (Eq.~(\ref{eq:vp_bayes_definition})) in these settings are also given by $v'_{\mathrm{C}}$ and $v'_{\mathrm{G}}$ in Eqs.~(\ref{eq:classical_bound}) and (\ref{eq:gaussian_bound}), because these bounds are independent of the measurement outcome $y$ (The classical bound $v'_{\mathrm{C}}$ matches with the result in Ref.~\cite{genoni2013optimal}, which is derived as the Bayesian RLD (right logarithmic derivative) quantum Cram\'{e}r-Rao bound. They also show the tightness of the bound.). This means that post-selection does not improve the estimation accuracy in these settings, which we will show is not true for non-Gaussian cases.
\section{Estimation using non-Gaussian states}\label{section:non_gaussian}
In this section, we show that the bounds introduced in Sec.~\ref{section:gaussian_bound} can be beaten using non-Gaussian states, and a heterodyne measurement, which is widely used for simultaneous measurements of both quadratures \cite{walker1986multiport,takeda2013generation,wodkiewicz1984operational,miyata2016implementation}. As examples of non-Gaussian states, we consider GKP state \cite{gottesman2001encoding}, which is expected to give high estimation accuracy as in Ref.~\cite{duivenvoorden2017single}, and Fock states, which are experimentally feasible for small photon numbers. 
\subsection{Heterodyne measurement}
\begin{figure}[ht]
\centering
\includegraphics{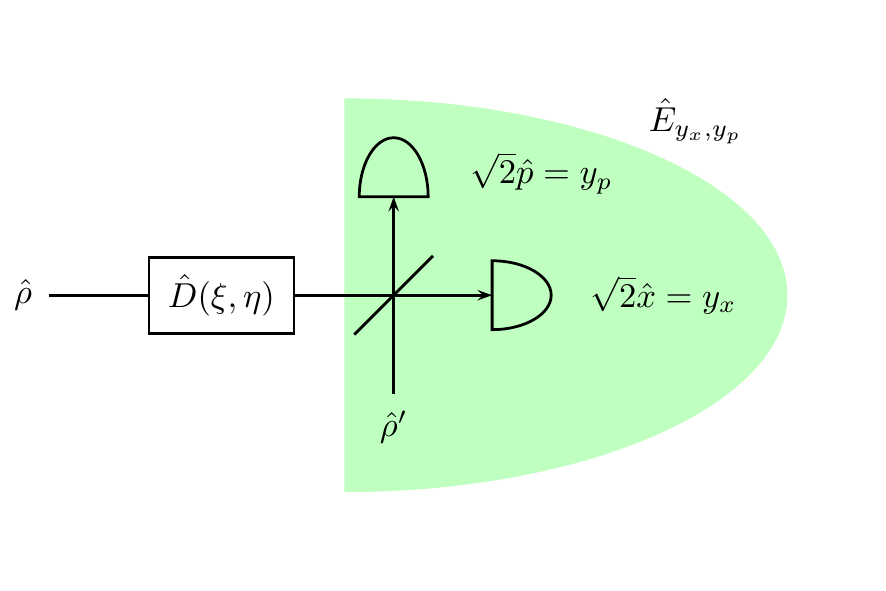}
\caption{The case when one uses heterodyne measurement as the measurement in Fig.~\ref{fig:displacement_estimation}. The state to be measured is first combined with an ancillary state by a half beam splitter, then $\sqrt{2}\hat{x}$ and $\sqrt{2}\hat{p}$ are measured on each mode. The corresponding POVM is expressed by Eq.~(\ref{eq:dualhomodyne_povm}).}
\label{fig:dual_homodyne}
\end{figure}
As the measurement $\{\Ehat_y\}$ for the system in Fig.~\ref{fig:displacement_estimation}, we consider a setup called heterodyne measurement shown in Fig.~\ref{fig:dual_homodyne}, which consists of a half beam splitter, an ancillary state $\rhohat'$, and two homodyne measurements. The input state after the displacement is combined with $\rhohat'$ by the half beam splitter, then $\sqrt{2}\hat{x}$ and $\sqrt{2}\hat{p}$ are measured on each mode. The case when $\rhohat'$ is a vacuum state can be seen as a sampling from Husimi Q function and has wide applications, \eg{} state verifications \cite{walker1986multiport,takeda2013generation}. Here we consider a more general case where any single-mode state can be used as $\rhohat'$ \cite{wodkiewicz1984operational,miyata2016implementation}. Denoting two measurement outcomes as $y=(\yx,\yp)$, the corresponding POVM element is \cite{leonhardt2010essential}
\begin{equation}
    \Ehat_{\yx,\yp}=\frac{1}{2\pi}\hat{D}(\yx,\yp)\rhohat'^*\hat{D}^\dagger(\yx,\yp),\label{eq:dualhomodyne_povm}
\end{equation}
where $\rhohat'^*$ is the operator whose matrix elements are the complex conjugate of those of $\rhohat'$. Because taking the complex conjugate corresponds to the time reversal operation, the Wigner function of $\rhohat'^*$ is given by $W_{\rhohat'^*}(x,p)=W_{\rhohat'}(x,-p)$.Therefore, substituting Eq.~(\ref{eq:dualhomodyne_povm}) into Eq.~(\ref{eq:cond_prob}), one can see that the conditional probability density $p(\yx,\yp|\xi,\eta)$ only depends on $(\xi-\yx,\eta-\yp)$:
\begin{equation}
    \begin{split}
    &p(\yx,\yp|\xi,\eta)\\
    =&\frac{1}{2\pi}\Tr{\hat{D}(\xi-\yx,\eta-\yp)\rhohat\hat{D}^\dagger(\xi-\yx,\eta-\yp)\rhohat'^*}\\
    =&f(\xi-\yx,\eta-\yp),
    \end{split}
\end{equation}
where the ``filter function'' $f$ is determined by $\rhohat$ and $\rhohat'$, as
\begin{equation}
    f(x,p):=\int W_{\rhohat}(x'-x,p'-p)W_{\rhohat'}(x',-p') dx'dp'\label{eq:filter_function}
\end{equation}
using Eq.~(\ref{eq:cond_prob_wigner}). From Eq.~(\ref{eq:post_prob}), the posterior distribution of $\xi,\eta$ is obtained by multiplying the prior distribution by the filter function displaced by the measurement outcome $\yx,\yp$:
\begin{equation}
    p(\xi,\eta|\yx,\yp)\propto p(\xi,\eta)f(\xi-\yx,\eta-\yp).\label{eq:bayesian_inference}
\end{equation}
Note that this form of measurement is not only experimentally feasible, but also quite general in the presence of post-selection, because having any POVM element $\Ehat$ is equivalent to taking $\rhohat'^*\propto \Ehat$ and post-selecting $\yx=\yp=0$ in this heterodyne setting.

\subsection{Estimation using GKP state}

GKP state, or grid state \cite{duivenvoorden2017single}, is a non-Gaussian state used for GKP error correction \cite{gottesman2001encoding}, and is defined as
\begin{equation}
    \ket{\mathrm{GKP}}\propto\sum_{s=-\infty}^\infty  \ket{x=\sqrt{2\pi}s}.
\end{equation}
Note that this state is not normalizable, therefore an unphysical state. The Wigner function of GKP state is a sum of Dirac delta functions:
\begin{equation}
    W_{\mathrm{GKP}}(x,p)\propto\sum_{s=-\infty}^\infty\sum_{t=-\infty}^\infty(-1)^{st}\delta(x-\sqrt{\pi/2}s)\delta(p-\sqrt{\pi/2}t)
\end{equation}
If one takes $\rhohat=\rhohat'=\ketbra{\mathrm{GKP}}{\mathrm{GKP}}$, the filter function Eq.~(\ref{eq:filter_function}) becomes
\begin{equation}
    f_{\mathrm{GKP}}(x,p)=\sum_{s=-\infty}^\infty\sum_{t=-\infty}^\infty\delta(x-\sqrt{2\pi}s)\delta(p-\sqrt{2\pi}t)
\end{equation}

\begin{figure}[htp]
    \centering
    \includegraphics[width=0.9\linewidth]{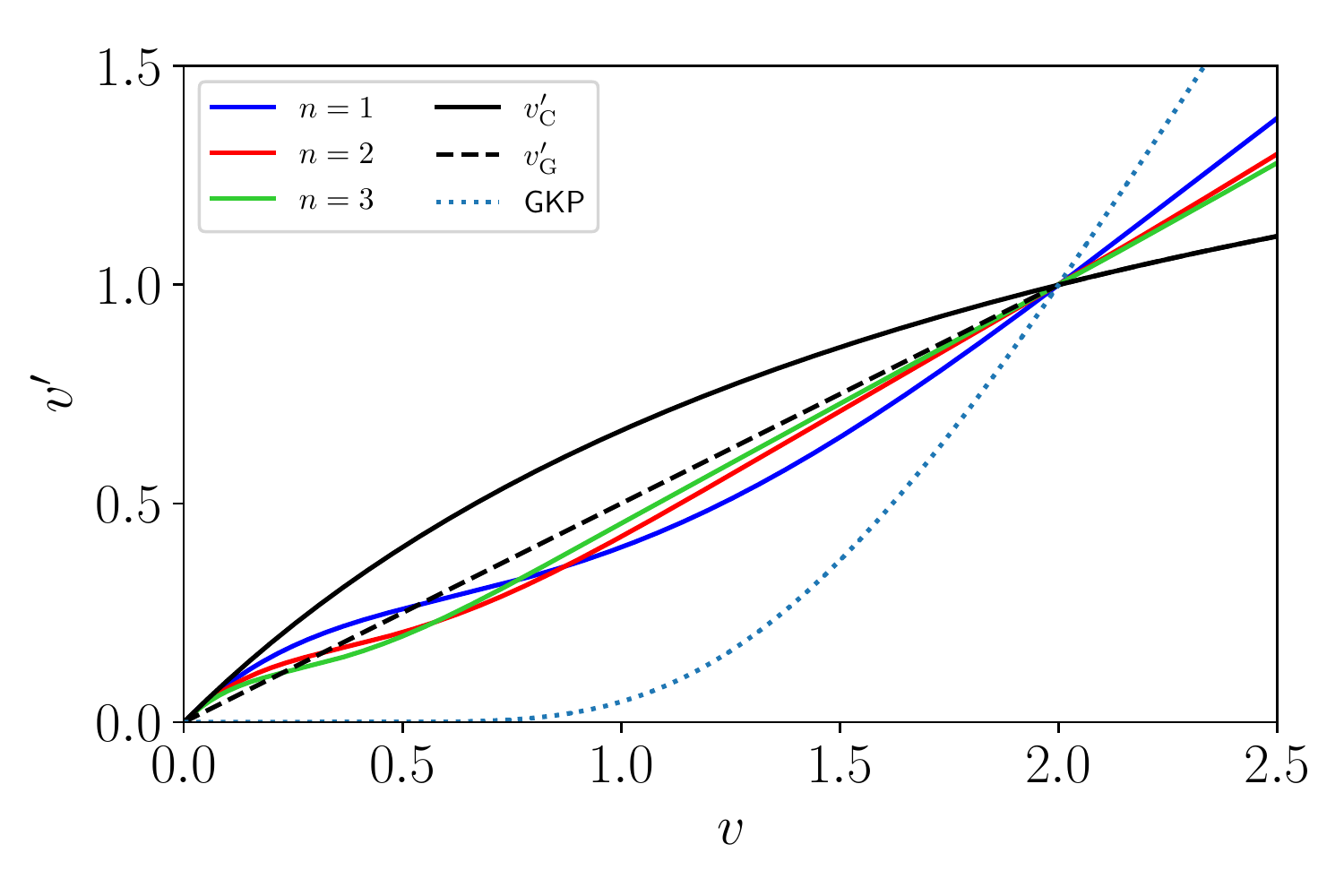}
    \caption{The relation between $v$ and $v'$. `$n=1$', `$n=2$', `$n=3$': when $\rhohat=\rhohat'=\ketbra{n}{n}$, and $(\yx,\yp)=(0,0)$ is post-selected. `GKP': when $\rhohat=\rhohat'=\ketbra{\mathrm{GKP}}{\mathrm{GKP}}$, and $(\yx,\yp)=(0,0)$ is post-selected. `$v'_{\mathrm{Bayes}} (n=1)$': when $\rhohat=\rhohat'=\ketbra{1}{1}$, and without post-selection. $v'_{\mathrm{C}}$ and $v'_{\mathrm{G}}$ are the classical and Gaussian bounds derived in Sec.~\ref{section:gaussian_bound}, respectively.}
    \label{fig:v_vs_vp}
\end{figure}

Reference \cite{duivenvoorden2017single} points out that this (unphysical) GKP state can achieve 0 estimation error in the non-Bayesian setting with the local unbiasedness condition. It is expected to achieve good estimation accuracy also in our setting, where we consider Bayesian estimation with post-selection. This is because the prior information that $(\xi,\eta)$ is near origin and the locally sharp structure of the filter function circumvent the uncertainty principle. The line labeled `GKP' in Fig.~\ref{fig:v_vs_vp} shows the relation between $v$ and $v'$, when one takes $\rhohat=\rhohat'=\ketbra{\mathrm{GKP}}{\mathrm{GKP}}$ and $(\yx,\yp)=(0,0)$ is post-selected. One can see that $v'$ is largely reduced compared to the Gaussian and classical bounds for small $v$. Note that even with the ideal GKP state, the Bayesian estimation error $v'$ remains finite due to the finite tail of the Gaussian prior distribution, unlike in frequentist's locally unbiased estimation schemes such as Ref.~\cite{duivenvoorden2017single}. On the other hand, $v'$ becomes larger than the classical bound in the region $v\geq 2$. This may be due to the fact that the filter function $f_{\mathrm{GKP}}(x,p)$ has a large variance. Note also the similarity of this setting with the error syndrome measurement for the GKP code \cite{gottesman2001encoding}, where the SUM gate is used instead of the beam splitter.

\subsection{Estimation using Fock states}
\begin{figure}[htp]
    \centering
    \includegraphics[width=0.9\linewidth]{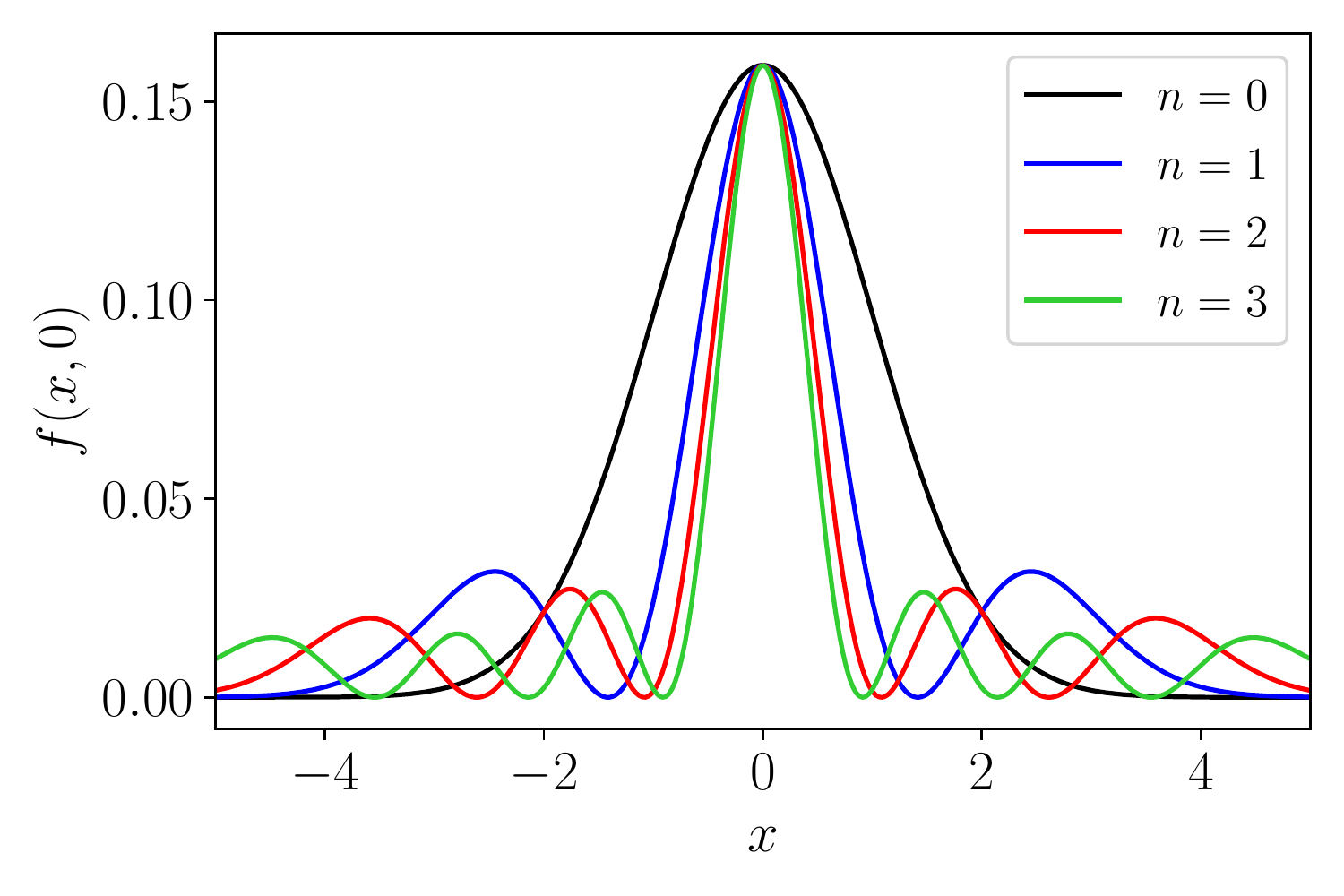}
    \caption{The filter function $f_n(x,0)$ when $\rhohat=\rhohat'=\ketbra{n}{n}$ for $ n=0,1,2,3$ (Eq.~(\ref{eq:fock_filter})).}
    \label{fig:fock_filters}
\end{figure}

While the GKP state is obviously useful for the estimation of displacement, experimental generation of (an approximation with physical state of) it has not been realized in optics. We now discuss whether the Gaussian bound can be beaten using experimentally feasible non-Gaussian states. Fock states $\ket{n}$ are the most simple examples of non-Gaussian states, and also experimentally feasible when $n$ is small \cite{ourjoumtsev2006quantum,yukawa2013generating}, although it is still harder than Gaussian states. If one takes $\rhohat=\rhohat'=\ketbra{n}{n}$, the filter function Eq.~(\ref{eq:filter_function}) becomes
\begin{equation}
    f_n(x,p)= \frac{1}{2\pi}\qty[L_n\qty(\frac{x^2+p^2}{2})]^2\exp(-\frac{x^2+p^2}{2}),\label{eq:fock_filter}
\end{equation}
where $L_n(\cdot)$ is the Laguerre polynomial. $f(x,0)$ for $n=0,1,2,3$ is shown in Fig.~\ref{fig:fock_filters}. The distance from the origin to the first zero of $f_n$ scales as $n^{-1/2}$ \cite{GATTESCHI20027}. Thus, roughly speaking, for sufficiently small $v$ and when $(\yx,\yp)=(0,0)$ is post-selected, $f_n$ acts like a Gaussian function of variance $\sim n^{-1}$, therefore Fock states are candidates for good input states. Note that when $\rhohat=\rhohat'^*$, $f_n(x,p)$ always has a maximum value $1/(2\pi)$ at the origin.

\begin{figure}[htp]
    \centering
    \includegraphics[width=0.9\linewidth]{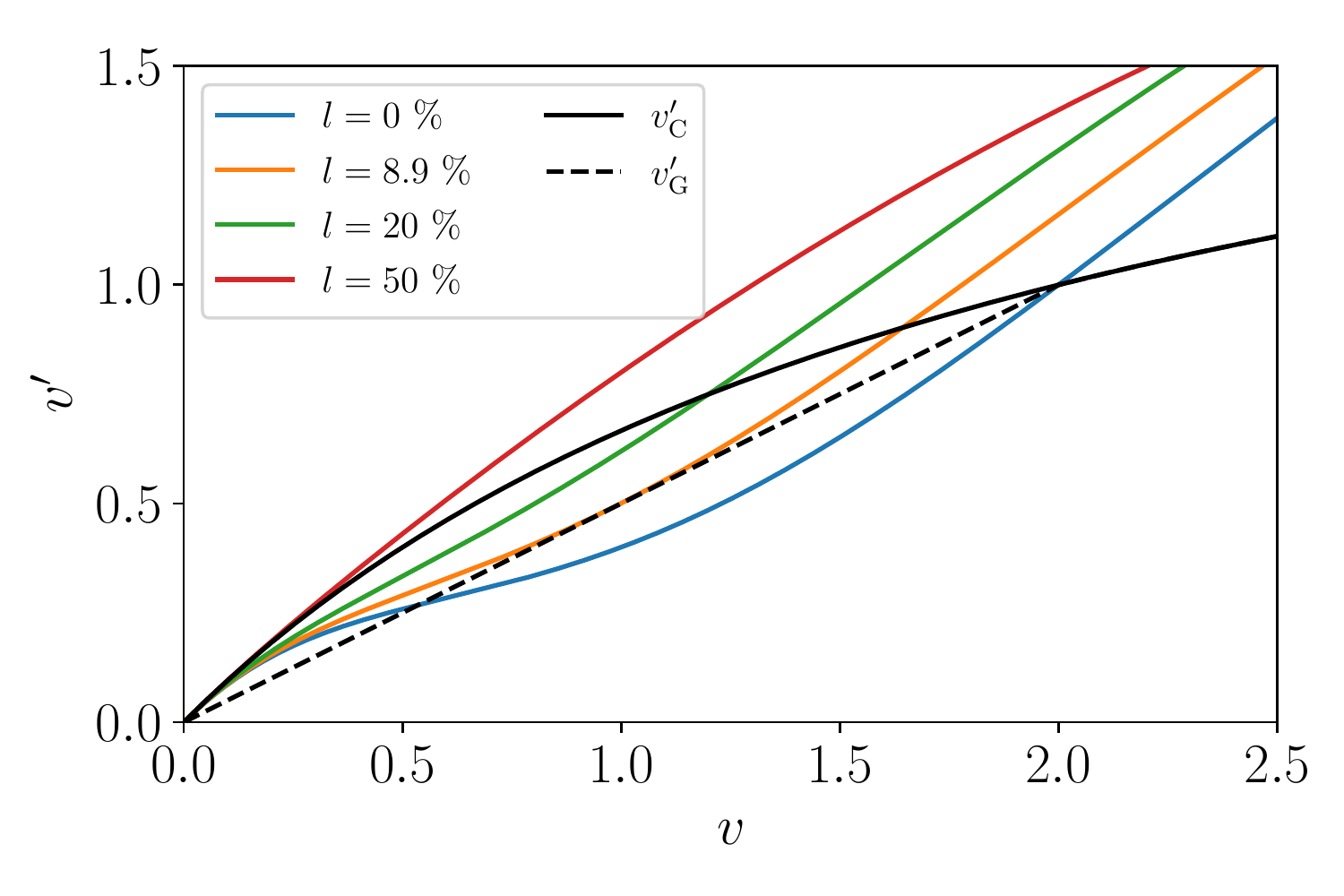}
    \caption{The relation between $v$ and $v'$, when $\rhohat=\rhohat'=(1-l)\ketbra{1}{1}+l\ketbra{0}{0}$, where $l$ corresponds to the photon loss rate.}
    \label{fig:loss_effect}
\end{figure}

Figure \ref{fig:v_vs_vp} shows the relation between $v$ and $v'$ for $n=1,2,3$.  For comparison, the Bayesian estimation error $v'_{\mathrm{Bayes}}$ without post-selection (Eq.~(\ref{eq:vp_bayes_definition})) for $n=1$ is also shown. One can see that the estimation accuracy is improved by post-selection, and even when $n=1$, the Gaussian bound is beaten in some range of $v$. We also calculate the effect of photon losses on both input states for the case $\rhohat=\rhohat'=\ketbra{1}{1}$, as is important for an actual experimental realization (Fig.~\ref{fig:loss_effect}). We assume the same amount of losses for $\rhohat$ and $\rhohat'$. Losses up to 8.9\% and 50\% are allowed for beating the Gaussian bound and the classical bound, respectively.

Note that practically post-selecting a single value of $(\yx,\yp)$ is impossible in continuous-variable measurements such as the heterodyne measurement, and one has to select events in some range of $(\yx,\yp)$ to have sufficiently high post-selection probability. In Appendix \ref{section:finite_prob}, we include an analysis of the case selecting finite range of outcomes with finite probability in the single-photon case. It shows that the classical and Gaussian bounds are still beaten in some range of $v$, with finite post-selection probability.

It is interesting to observe that for all GKP state and Fock states cases, $v'$ becomes $1$ at $v=2$. In fact, one can explicitly show that when $v=2$, $v'\geq 1$ holds for arbitrary $\rhohat$ and $\Ehat_y$, and the equality holds when $\rhohat\propto\Ehat_y\propto\ketbra{\psi}{\psi}$ for some $\ket{\psi}$. See Appendix \ref{section:vp_limit_v2} for the derivation.

\section{Analysis on lower bounds for non-Gaussian states}\label{section:lower_bound}
We return to the general setting in Sec.~\ref{section:estimation_problem}. One natural matter of interest is to find better input states, and find the limit of estimation using non-Gaussian states. Generally, preparation of non-Gaussian states with large photon number is difficult, and experimentally feasible states lie in a subspace with some maximum photon number \cite{yukawa2013generating}. In this section, we discuss lower bounds on $v'$, when $\rhohat$ is a superposition of Fock states up to the maximum photon number $n$.

In quantum estimation theory, the Cram\'{e}r-Rao bound \cite{braunstein1994statistical,paris2009quantum,martinez2017quantum,cramer1999mathematical,helstrom1976quantum} is often used to obtain a lower bound on the estimation error. Because we have prior information and post-select a single measurement outcome in our case, we use an inequality called Ghosh's bound instead \cite{ghosh1993cramer}. See Appendix~\ref{section:gen_van_trees} for the detailed derivation of Ghosh's bound and its evaluation. According to the inequality, a lower bound for $v'$, when the measurement outcome is $y$, is given by
\begin{align}
    v'\geq\frac{4}{F_0(y)+F(y)},\label{eq:gen_van_trees}
\end{align}
where $F_0(y)$ and $F(y)$ are defined as:
\begin{align}
    F_0(y)&:=\sum_i\int p(\btheta|y)\qty(-\frac{\partial^2}{\partial \theta_i^2} \log p(\btheta)) d\btheta,\label{eq:fisher0}\\
    F(y)&:=\sum_i\int p(\btheta|y)\qty(-\frac{\partial^2}{\partial \theta_i^2} \log p(y|\btheta)) d\btheta.\label{eq:fisher}
\end{align}
Here we write $\btheta:=(\theta_1,\theta_2):=(\xi,\eta)$, $(\qhat_1,\qhat_2):=(\hat{x},\hat{p})$, and $d\btheta:=d\theta_1 d\theta_2$. The quantities $F(y)$ and $F_0(y)$ can be considered as variants of Fisher information. There is an upper bound on $F(y)$:
\begin{equation}
    F(y) \leq \sum_i\int d\btheta \frac{p(\btheta)}{p(y)}\Tr{\qty(2\qhati\rhohat \qhati+\qhati^2 \rhohat+\rhohat \qhati^2)\Ehatyth},\label{eq:fisher_expression}
\end{equation}
where
\begin{equation}
    \Ehatyth:=\hat{D}^\dagger(\btheta)\Ehat_y\hat{D}(\btheta)
\end{equation}
(see Appendix~\ref{section:gen_van_trees} for the derivation). For the Gaussian prior distribution Eq.~(\ref{eq:gaussian_prior_dist}), $F_0(y)$ is simply a constant:
\begin{equation}
    F_0(y)=\frac{4}{v}.
\end{equation}
Therefore, inequality Eq.~(\ref{eq:gen_van_trees}) can be transformed as
\begin{equation}
    \frac{1}{v'}\leq\frac{1}{v}+\frac{F(y)}{4}.\label{eq:vp_bound}
\end{equation}
An upper bound for $F(y)$, in the case when the maximum photon number of $\rhohat$ is $n$, can be obtained using the Schwarz inequality. For pure $\rhohat=\ketbra{\psi}{\psi}$, we have
\begin{equation}
    \begin{split}
        \sum_i \Tr{\qhati\rhohat \qhati \Ehatyth}&\leq \sqrt{\sum_i\Tr{\qhati^2\rhohat \qhati^2\rhohat}}\sqrt{2\Tr\Ehat_y^2}\\
        &=\sqrt{\sum_i(\Braket{\psi|\qhati^2|\psi})^2}\sqrt{2\Tr\Ehat_y^2}\\
        &\leq \sum_i \Braket{\psi|\qhati^2|\psi}\sqrt{2\Tr\Ehat_y^2}\\
        &=\Braket{\psi|(2\hat{n}+1)|\psi}\sqrt{2\Tr\Ehat_y^2}\\
        &\leq \sqrt{2}(2n+1)\sqrt{\Tr\Ehat_y^2},
    \end{split}\label{eq:upper_bound_1}
\end{equation}
Here we used the Schwarz inequality with respect to the inner product $\left<A,B\right>:=\sum_i \Tr{\hat{A}_i \hat{B}_i^\dagger}$ in the first inequality, taking $\hat{A}_i=\hat{q}_i\hat{\rho}\hat{q}_i, \hat{B}_i=\Ehatyth$. The second inequality follows from the fact that $x^2+y^2\leq(x+y)^2$ for $x,y>0$. In the same way, considering an inner product $\left<A,B\right>:=\Tr{\hat{A} \hat{B}^\dagger}$ and taking $\hat{A}=\sum_i \hat{q}_i^2\hat{\rho}$ and $\hat{B}=\Ehatyth$, we get 
\begin{equation}
    \begin{split}
        \sum_i \Tr{\qhati^2\rhohat \Ehatyth}&\leq \sqrt{\Tr{(\sum_i\qhati^2)^2\rhohat^2}}\sqrt{\Tr\Ehat_y^2}\\
        &\leq \sqrt{\bra{\psi}(\sum_i\qhati^2)^2\ket{\psi}}\sqrt{\Tr\Ehat_y^2}\\
        &=\sqrt{\bra{\psi}(2\hat{n}+1)^2\ket{\psi}}\sqrt{\Tr\Ehat_y^2}\\
        &\leq (2n+1)\sqrt{\Tr\Ehat_y^2}.
    \end{split}\label{eq:upper_bound_2}
\end{equation}
Therefore, from Eq.~(\ref{eq:fisher_expression}), we obtain
\begin{equation}
    F(y)\leq 2(\sqrt{2}+1)(2n+1)\frac{\sqrt{\Tr\Ehat_y^2}}{p(y)}.\label{eq:fisher_info_bound}
\end{equation}
Since the left-hand sides of Eqs.~(\ref{eq:upper_bound_1}) and (\ref{eq:upper_bound_2}) are linear in $\rhohat$, the bound Eq.~(\ref{eq:fisher_info_bound}) also holds for arbitrary mixed $\rhohat$. From Eqs.~(\ref{eq:fisher_info_bound}) and (\ref{eq:vp_bound}), one obtains a lower bound of $v'$:
\begin{equation}
    \frac{1}{v'}\leq\frac{1}{v}+ (\sqrt{2}+1)(n+1/2)\frac{\sqrt{\Tr\Ehat_y^2}}{p(y)}.\label{eq:vp_bound_final}
\end{equation}

\begin{figure}[htp]
    \centering
    \includegraphics[width=0.9\linewidth]{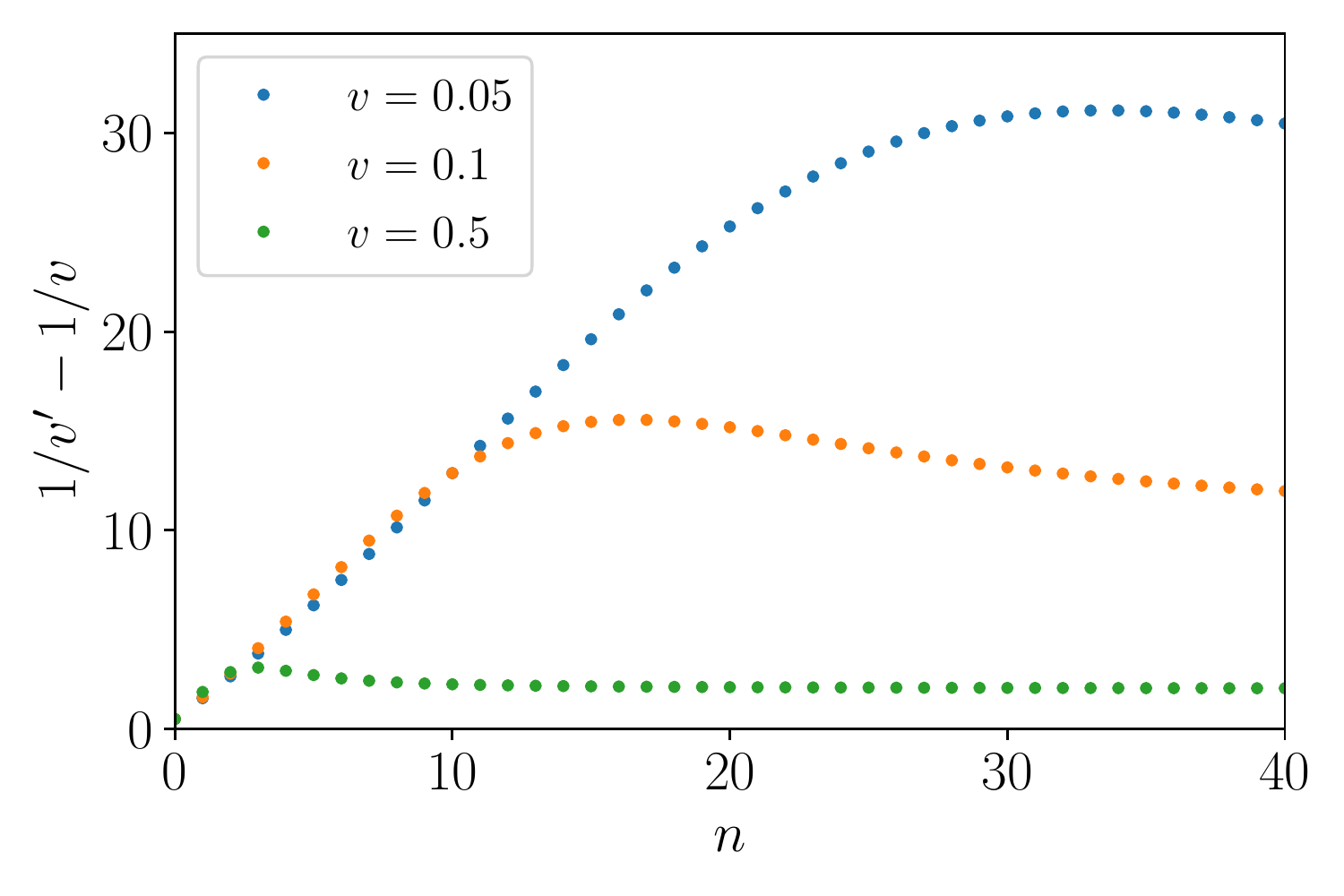}
    \caption{The relation between $n$ and $1/v'-1/v$ for fixed values of $v$, when $\rhohat=\Ehat_y=\ketbra{n}{n}$.}
    \label{fig:n_vs_fisher}
\end{figure}

\begin{figure}[htp]
    \centering
    \includegraphics[width=0.9\linewidth]{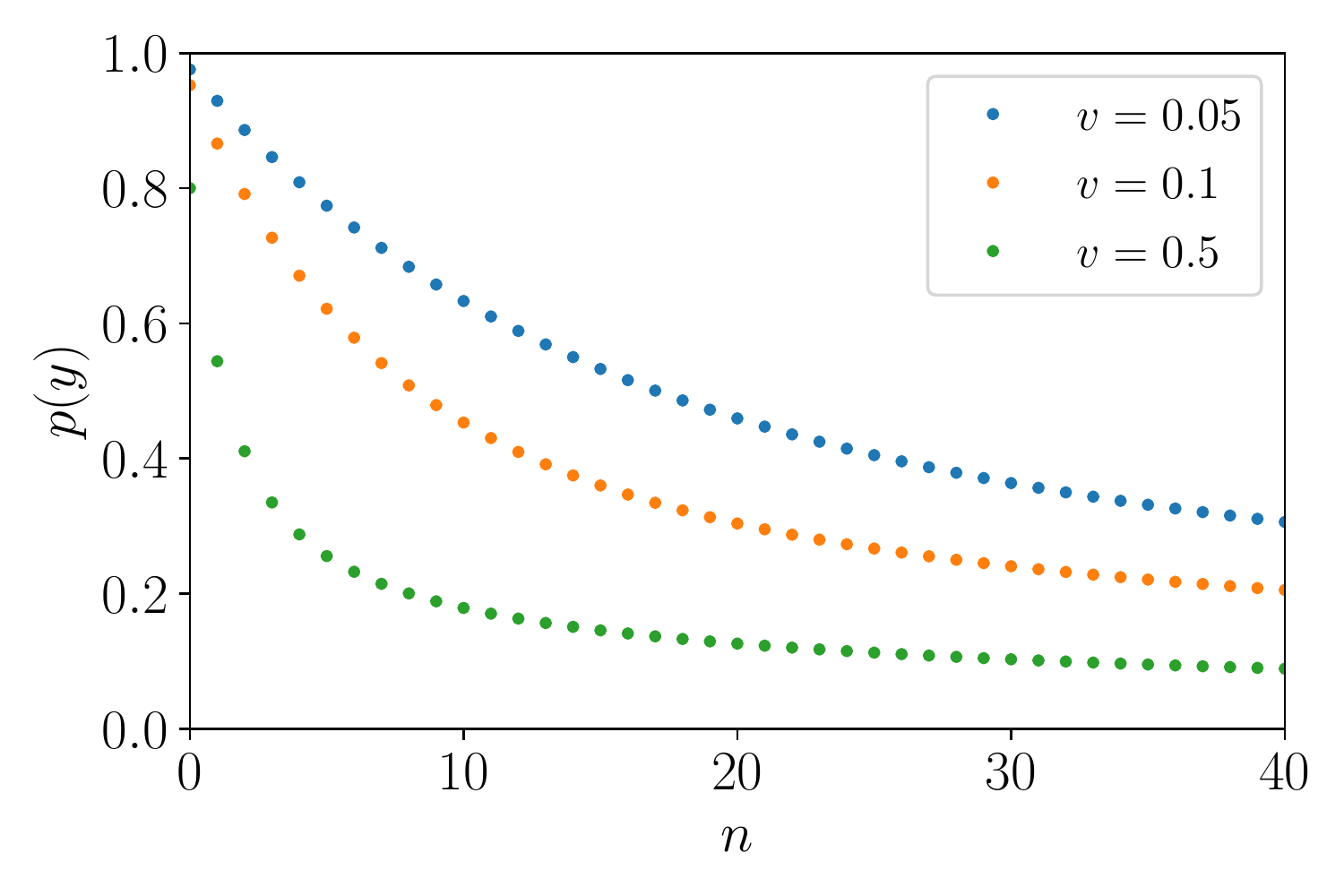}
    \caption{The relation between $n$ and $p(y)$ for fixed values of $v$, when $\rhohat=\Ehat_y=\ketbra{n}{n}$.}
    \label{fig:n_vs_py}
\end{figure}

The $\mathcal{O}(n)$ upper bound of $F(y)$ is similar to the result in Ref.~\cite{duivenvoorden2017single}, which studies the case with no prior information and derives an $\mathcal{O}(n)$ upper bound for the Fisher information. They stated that whether the $n$ scaling of the Fisher information is achievable or not is an open problem, and argued that Fock states $\ket{n}$ do not achieve it due to the fringe of the likelihood function (the filter function $f_n(x,p)$ in our setting) \cite{duivenvoorden2017single}. Similarly in our problem, if we take $\rhohat=\Ehat_y=\ketbra{n}{n}$ and increase $n$ for fixed $v$, $1/v'-1/v$ only behaves $\sim n$ up to some maximum $n$ (which depends on $v$) and then decreases, while $p(y)$ also gradually decreases (Figs.~\ref{fig:n_vs_fisher} and \ref{fig:n_vs_py}). This result is consistent with the observation that the filter function Eq.~(\ref{eq:fock_filter}) behaves like a Gaussian with variance $\sim n^{-1}$ only when $v$ is smaller than this variance.

The bound Eq.~(\ref{eq:vp_bound_final}) also depends on the normalized post-selection probability density $\frac{p(y)}{\sqrt{\Tr\Ehat_y^2}}$. 
This implies that one can possibly obtain large $F(y)$ at the expense of small post-selection probability density. In fact, even when $n=1$, one can construct an example where $1/v'-1/v$ is inversely proportional to $\frac{p(y)}{\sqrt{\Tr\Ehat_y^2}}$ when $\frac{p(y)}{\sqrt{\Tr\Ehat_y^2}}\to 0$. See Appendix \ref{section:smallp} for a detail.

\section{Conclusion}\label{sec:conclusion}
We have considered the problem of estimating the amount of random displacement that follows an isotropic Gaussian distribution. We have shown that there is a lower bound on the estimation error $v'$ when only Gaussian states and Gaussian operations are used. This bound can be beaten for some range of the prior variance $v$, using only linear optics and simple non-Gaussian states such as single-photon states. When the maximum photon number of the input state is $n$, $1/v'-1/v$ has an $\mathcal{O}(n)$ upper bound which is also inversely proportional to the post-selection probability.

Because of the similarity of the estimation of displacement to the error syndrome measurement in GKP code \cite{gottesman2001encoding}, extending the method proposed in this paper may lead to an error correcting code using experimentally feasible non-Gaussian states such as single photon states, which is important from a practical point of view. The Gaussian bound derived in this paper seems to correspond to the impossibility of correcting Gaussian errors using Gaussian states \cite{niset2009no}. Finding more direct connection between the estimation of displacement and the quantum error correction is a promising future work. 

Another possible future work is to find the best input state and POVM for a given constraint, \eg{} maximum photon number. The bound Eq.~(\ref{eq:vp_bound_final}) is not necessarily tight, but for quantum estimation of displacement, many kinds of lower bounds for the estimation error are known \cite{tsang2020physics,nagaoka2005new,rubio2020bayesian}, and some of them have been shown to be efficiently solvable \cite{conlon2020efficient,sidhu2019tight}. It is possible that they can be extended to our setting.
\begin{acknowledgments}
This work was partly supported by JST [Moonshot R\&D][Grant No.~JPMJMS2064], JSPS KAKENHI (Grant No.~18H05207, No.~21J11615), UTokyo Foundation, and donations from Nichia Corporation. 
\end{acknowledgments}

\appendix
\section{Derivation of Ghosh's bound and an upper bound of Fisher information}\label{section:gen_van_trees}
We consider the estimation problem described in Sec.~\ref{section:estimation_problem}. We write $\xi,\eta$ as $\theta_1,\theta_2$, and $\hat{x},\hat{p}$ as $\qhat_1,\qhat_2$ in this section, for convenience. We also just write $(\theta_1,\theta_2)$ as $\btheta$, and $d\theta_1 d\theta_2$ as $d\btheta$.

Let us denote the estimated value of $\theta_i$ by $\tilde{\theta}_i(y)$. The Van Trees inequality \cite{van1968detection,gill1995applications,paris2009quantum}, also called Bayesian Cram\'{e}r-Rao inequality gives a lower bound for the Bayesian estimation error. In a weaker form than the original matrix inequality, it can be written as
\begin{equation}
    \sum_i\int dy\int d\btheta \qty(\tilde{\theta}_i(y)-\theta_i)^2 p(y,\btheta) \geq\frac{4}{F_{0}^{(VT)}+F^{(VT)}},\label{eq:van_trees}
\end{equation}
where
\begin{equation}
    \begin{split}
    F_{0}^{(VT)}&:=\sum_i\int d\btheta p(\btheta)\qty(-\frac{\partial^2}{\partial \theta_i^2} \log p(\btheta)),\\
    F^{(VT)}&:=\sum_i\int dy\int d\btheta p(y,\btheta)\qty(-\frac{\partial^2}{\partial \theta_i^2} \log p(y|\btheta)).
    \end{split}
\end{equation}

$F_{0}^{(VT)}$ is the \apriori{} Fisher information of the prior distribution $p(\theta)$, and $F^{(VT)}$ is the expectation value of the Fisher information with respect to the prior distribution of $\theta$, which can be considered as the average information obtained by the measurement.
Eq.~(\ref{eq:van_trees}) can be used to evaluate the mean square error of the estimator, averaged over the measurement outcome $y$ and the parameter $\theta$. However, the Van Trees inequality cannot be directly applied to our case, because we allow post-selection of $y$. To generalize the Van Trees inequality for the case including post-selection, consider the following equation
\begin{equation}
    \sum_i\int (\tilde{\theta}_i(y)-\theta_i)\frac{\partial p(\btheta|y)}{\partial \theta_i}d\btheta=2,\label{eq:partial_integration}
\end{equation}
which can be shown using a partial integration. Using the Schwarz inequality, one obtains
\begin{equation}
    \sum_i\int \qty(\tilde{\theta}_i(y)-\theta_i)^2 p(\btheta|y)d\btheta \geq\frac{4}{\tilde{F}(y)},\label{eq:gen_van_trees_appendix}
\end{equation}
where $\tilde{F}(y)$ is
\begin{equation}
\begin{split}
    \tilde{F}(y)&:=\sum_i\int\frac{1}{p(\btheta|y)}\qty(\frac{\partial p(\theta|y)}{\partial \theta_i})^2 d\btheta\\
    &=\sum_i\int\qty(\frac{\partial p(\btheta|y)}{\partial \theta_i})\qty(\frac{\partial }{\partial \theta_i} \log p(\btheta|y)) d\btheta\\
    &=\sum_i\int p(\btheta|y)\qty(-\frac{\partial^2}{\partial \theta_i^2} \log p(\btheta|y)) d\btheta.
\end{split}
\end{equation}

Equation (\ref{eq:gen_van_trees_appendix}) is called Ghosh's bound \cite{ghosh1993cramer} in statistics.

$\tilde{F}(y)$ can be further decomposed as
\begin{align}
    \tilde{F}(y)&=F_0(y)+F(y),\label{eq:fisher_decomposition}\\
    F_0(y)&:=\sum_i\int p(\btheta|y)\qty(-\frac{\partial^2}{\partial \theta_i^2} \log p(\btheta)) d\btheta,\label{eq:fisher0_appendix}\\
    F(y)&:=\sum_i\int p(\btheta|y)\qty(-\frac{\partial^2}{\partial \theta_i^2} \log p(y|\btheta)) d\btheta.\label{eq:fisher_appendix}
\end{align}
Equation (\ref{eq:gen_van_trees_appendix}), together with Eqs.~(\ref{eq:fisher_decomposition}), (\ref{eq:fisher0_appendix}), and (\ref{eq:fisher_appendix}) gives the variant of Van Trees inequality with post-selection. $F_{0}^{(VT)}$ and $F^{(VT)}$ can be obtained as the expectation values of $F_0(y)$ and $F(y)$ with respect to $p(y)$. Therefore, if one takes the expectation values of both sides of Eq.~(\ref{eq:gen_van_trees_appendix}) with respect to $p(y)$, and uses the concavity of the function $1/t$, one obtains
\begin{equation}
   \sum_i\int dy\int d\btheta \qty(\tilde{\theta}_i(y)-\theta_i)^2 p(y,\btheta) \geq\frac{4}{\int p(y)\tilde{F}(y)dy},
\end{equation}
which reproduces Eq.~(\ref{eq:van_trees}). To find an expression for $F(y)$ in our case, one transforms Eq.~(\ref{eq:fisher_appendix}) as
\begin{equation}
    \begin{split}
        F(y)&=\sum_i\int p(\btheta|y)\qty(-\frac{\partial^2}{\partial \theta_i^2} \log p(y|\btheta)) d\btheta\\
        &=\sum_i\int\frac{p(\btheta)}{p(y)}\qty(\frac{1}{p(y|\btheta)}\qty(\frac{\partial p(y|\btheta)}{\partial \theta_i})^2-\frac{\partial^2p(y|\btheta)}{\partial \theta_i^2}) d\btheta.\label{eq:fisher_another_form}
    \end{split}
\end{equation}
Using that from Eq.~(\ref{eq:cond_prob}),
\begin{equation}
    p(y|\btheta)=\Tr{\rhohat\Ehatyth},
\end{equation}
where
\begin{equation}
    \Ehatyth:=\hat{D}^\dagger(\btheta)\Ehat_y\hat{D}(\btheta),
\end{equation}
one can explicitly calculate terms in Eq.~(\ref{eq:fisher_another_form}) as
\begin{align}
    \sum_i\qty(\frac{\partial}{\partial \theta_i}\Tr{\rhohat\Ehatyth})^2&=-\sum_i\Tr{\qty(\qhati\rhohat-\rhohat\qhati)\Ehatyth}^2,\\
    \sum_i\frac{\partial^2}{\partial \theta_i^2}\Tr{\rhohat\Ehatyth}&=\sum_i\Tr{\qty(2\qhati\rhohat\qhati-\qhati^2\rhohat-\rhohat\qhati^2)\Ehatyth},
\end{align}
resulting in
\begin{equation}
    \begin{split}
        F(y) &= \sum_i\int d\btheta \frac{p(\btheta)}{p(y)}\Biggl[-\frac{\Tr{\qty(\qhati\rhohat-\rhohat\qhati)\Ehatyth}^2}{\Tr{\rhohat\Ehatyth}}\\
        &\quad\quad+\Tr{\qty(-2\qhati\rhohat \qhati+\qhati^2 \rhohat+\rhohat \qhati^2)\Ehatyth}\Biggr].
    \end{split}
\end{equation}
An upper bound of $F(y)$, with a slightly simpler form can be obtained by applying the Schwarz inequality to
\begin{equation}
    \Tr{\qhati\rhohat\Ehatyth}=\Tr{\qty(\Ehatyth^{1/2}\qhati\rhohat^{1/2})\cdot\qty(\rhohat^{1/2}\Ehatyth^{1/2})},
\end{equation}
obtaining
\begin{equation}
    \begin{split}
        -\frac{\Tr{\qty(\qhati\rhohat-\rhohat\qhati)\Ehatyth}^2}{\Tr{\rhohat\Ehatyth}}&\leq\frac{4\qty|\Tr{\qhati\rhohat\Ehatyth}|^2}{\Tr{\rhohat\Ehatyth}}\\
        &\leq 4\Tr{\qhati\rhohat\qhati\Ehatyth},
    \end{split}
\end{equation}
which leads to
\begin{equation}
    F(y) \leq \sum_i\int d\btheta \frac{p(\btheta)}{p(y)}\Tr{\qty(2\qhati\rhohat \qhati+\qhati^2 \rhohat+\rhohat \qhati^2)\Ehatyth}.\label{eq:fisher_bound_1}
\end{equation}
\section{Derivation of Eq.~(\ref{eq:det_ineq})}\label{section:det_ineq}
Suppose $A,B$ are positive definite $2\times 2$ matrices. Then, we have
\begin{equation}
    \begin{split}
    \det(A+B)&=\det A \det(I+A^{-1/2}BA^{-1/2}),
    \end{split}\label{eq:det_ineq_deriv_1}
\end{equation}
where $I$ is the identity matrix. Since $A^{-1/2}BA^{-1/2}$ is positive definite and $\det(A^{-1/2}BA^{-1/2})=\frac{\det B}{\det A}$, we can denote 2 eigenvalues of $A^{-1/2}BA^{-1/2}$ as $x$ and $\frac{1}{x}\frac{\det B}{\det A}$. Then, we get
\begin{equation}
    \begin{split}
        \det(I+A^{-1/2}BA^{-1/2})&=(1+x)\qty(1+\frac{1}{x}\frac{\det B}{\det A})\\
        &=1+\frac{\det B}{\det A}+x+\frac{1}{x}\frac{\det B}{\det A}\\
        &\geq 1+\frac{\det B}{\det A}+2\sqrt{\frac{\det B}{\det A}}.
    \end{split}\label{eq:det_ineq_deriv_2}
\end{equation}

From Eqs.~(\ref{eq:det_ineq_deriv_1}) and (\ref{eq:det_ineq_deriv_2}), Eq.~(\ref{eq:det_ineq}) is obtained.

\section{Analysis of the case with a finite post-selection probability}\label{section:finite_prob}
\begin{figure}[htp]
    \centering
    \includegraphics[width=0.9\linewidth]{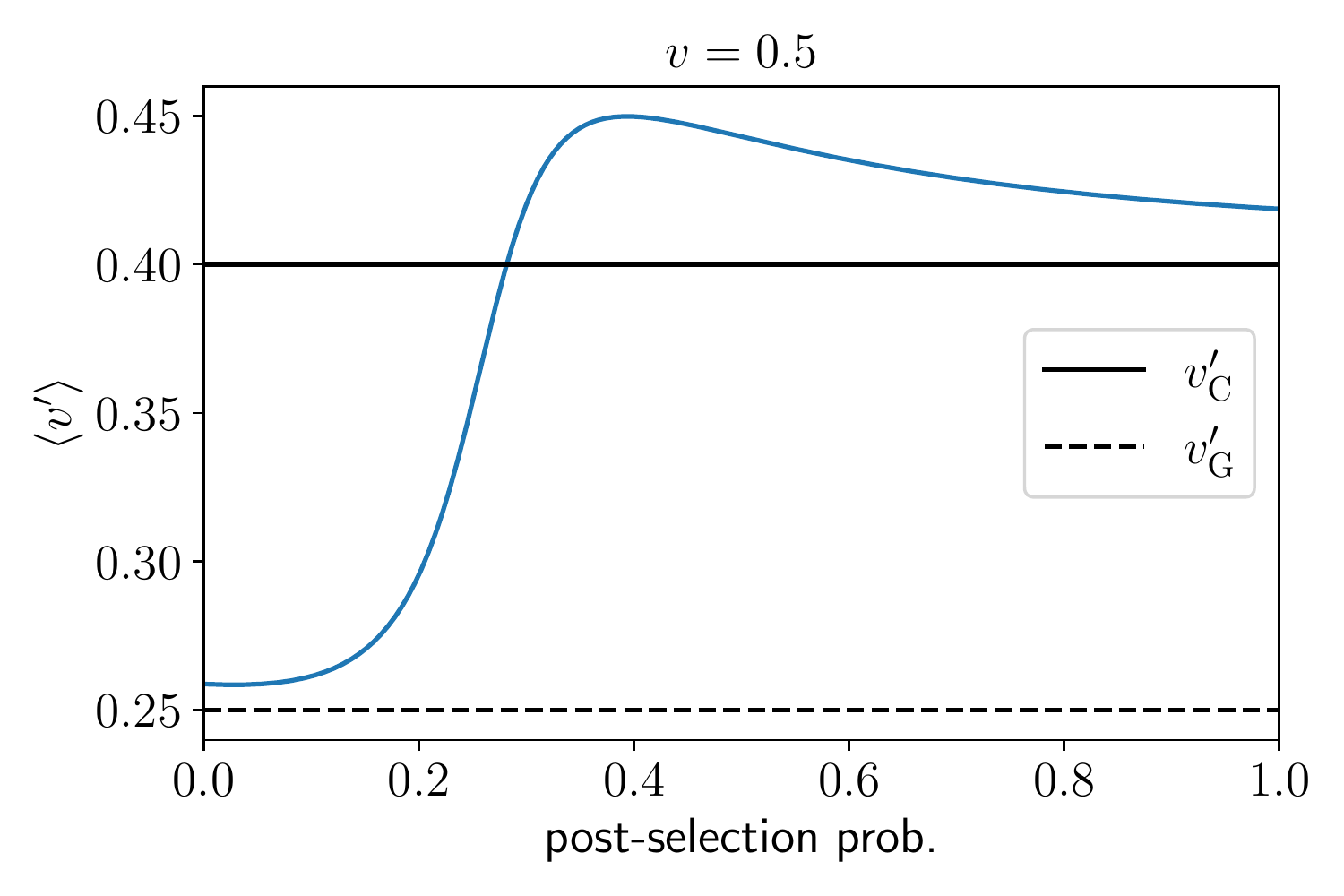}
    \includegraphics[width=0.9\linewidth]{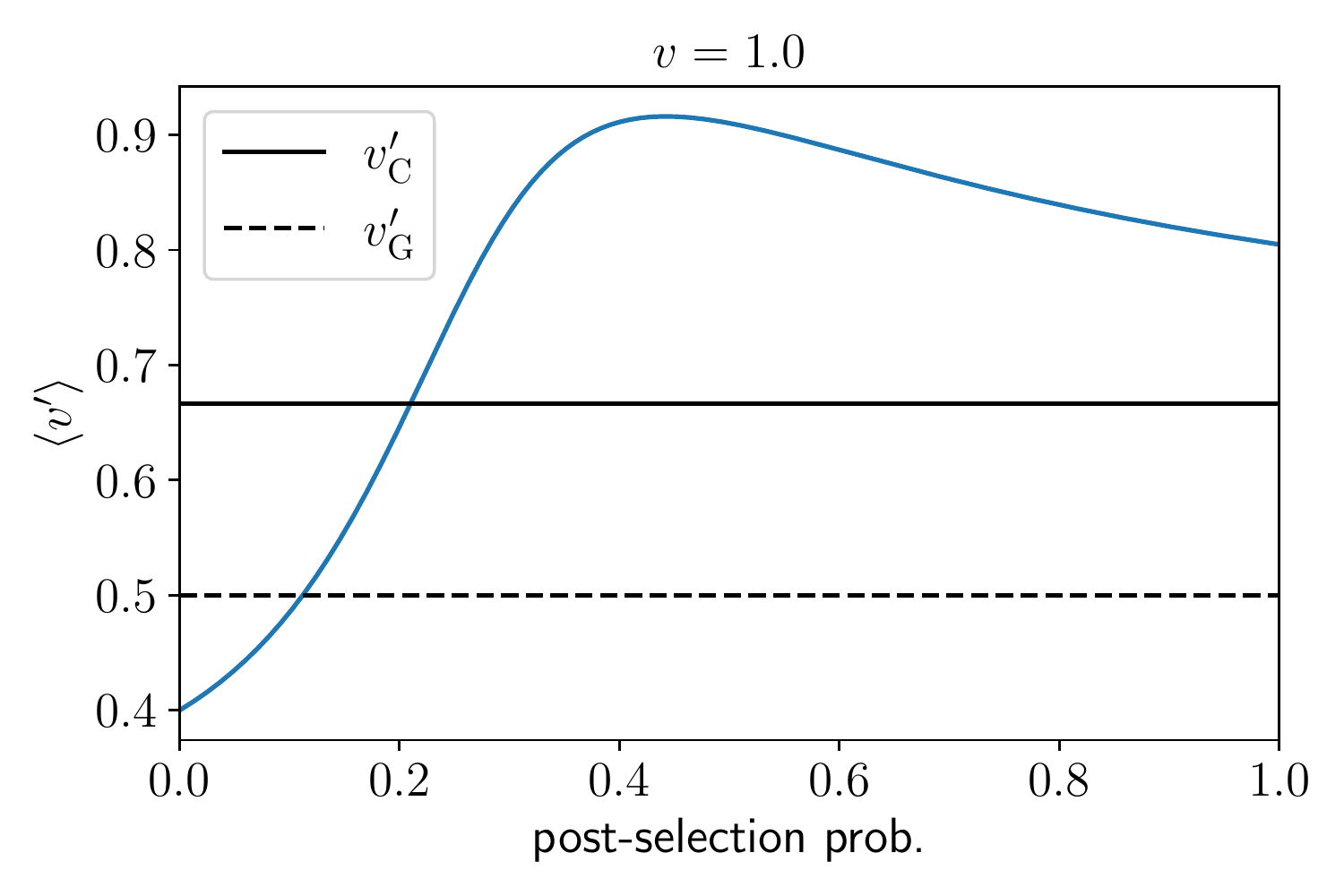}
    \includegraphics[width=0.9\linewidth]{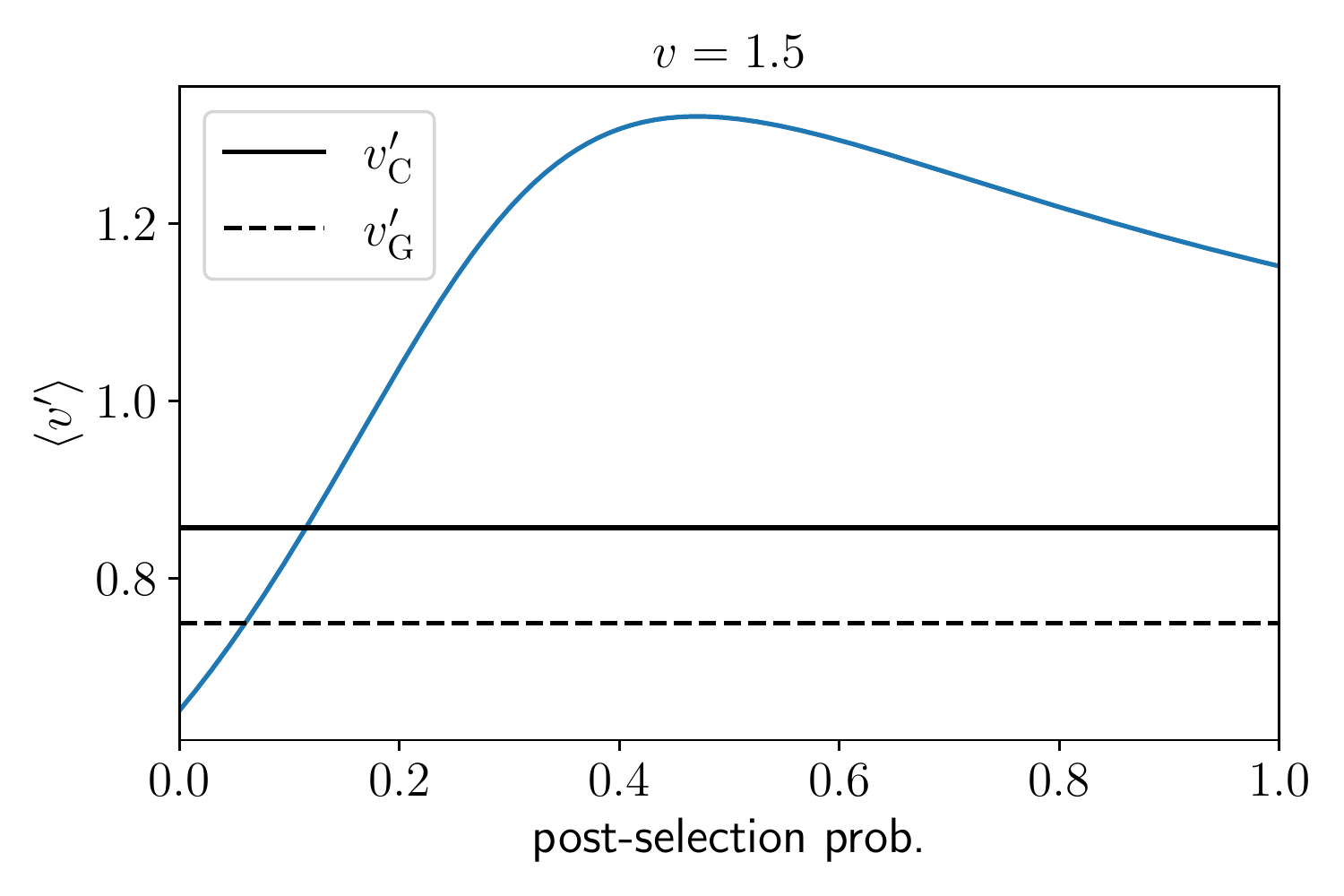}
    \caption{The relation between the post-selection probability (Eq.~(\ref{eq:post_select_prob})) and $\left<v'\right>$ (Eq.~(\ref{eq:vp_ave})) for $v=0.5,1.0,1.5$, when we take $\rhohat=\rhohat'=\ketbra{1}{1}$ in the setup in Sec.~\ref{section:non_gaussian} and post-select a range $\yx^2+\yp^2\leq r^2$ for $r>0$.}
    \label{fig:finite_prob}
\end{figure}

We consider the setup in Sec.~\ref{section:non_gaussian}, and take $\rhohat=\rhohat'=\ketbra{1}{1}$. Rather than post-selecting a single point $(\yx,\yp)=(0,0)$, we post-select a range $\yx^2+\yp^2\leq r^2$ for $r>0$. By increasing $r\to\infty$, the post-selection probability converges to $1$. We define the estimation error $\left<v'\right>$ as the avarage of $v'$ within the selected range of $y$:
\begin{equation}
    \left<v'\right>=\frac{\int_{\yx^2+\yp^2\leq r^2} v'(\yx,\yp)p(\yx,\yp) d\yx d\yp}{p(\yx^2+\yp^2\leq r^2)},\label{eq:vp_ave}
\end{equation}
where $p(\yx^2+\yp^2\leq r^2)$ is the post-selection probability:
\begin{equation}
    p(\yx^2+\yp^2\leq r^2)=\int_{\yx^2+\yp^2\leq r^2} p(\yx,\yp) d\yx d\yp\label{eq:post_select_prob}
\end{equation}
In Fig.~\ref{fig:finite_prob}, the relation between the post-selection probabilty and $\left<v'\right>$ is plotted, for $v=0.5,1.0,1.5$ (See also Fig.~\ref{fig:v_vs_vp}). It can be seen that the classical and Gaussian bounds are still beaten in some range of $v$, with finite post-selection probability.

\section{Proof that $v'\geq 1$ when $v=2$ for arbitrary states and measurements}\label{section:vp_limit_v2}
Because the convolution of two Wigner functions can be obtained by considering half-beamsplitter interaction and tracing out one of the modes, the conditional probability Eq.~(\ref{eq:cond_prob_wigner}) can be expressed as
\begin{equation}
    \begin{split}
    p(y|\xi,\eta)&=2\pi\int W_{\rhohat}(x-\xi/2,p-\eta/2)W_{\Ehat_y}(x+\xi/2,p+\eta/2) dxdp\\
    &=\pi\int W_{\rhohat}\qty(\frac{x-\xi/\sqrt{2}}{\sqrt{2}},\frac{p-\eta/\sqrt{2}}{\sqrt{2}})\\
    &\quad \quad W_{\Ehat_y}\qty(\frac{x+\xi/\sqrt{2}}{\sqrt{2}},\frac{p+\eta/\sqrt{2}}{\sqrt{2}}) dxdp\\
    &=\pi W_{\hat{\sigma}}(\xi/\sqrt{2},\eta/\sqrt{2}),
    \end{split}
\end{equation}
where $\hat\sigma$ is a positive operator defined as
\begin{equation}
    \hat\sigma := \Tr_1[\hat{B}^\dagger(\rhohat\otimes\Ehat_y) \hat{B}].
\end{equation}
Here $\Tr_1$ denotes the partial trace of the first system, and the half-beamsplitter operator $\hat{B}$ acts as
\begin{equation}
    \begin{split}
    \hat{B}^\dagger \hat{x}_1 \hat{B}&=\frac{1}{\sqrt{2}}(\hat{x}_1+\hat{x}_2),\hat{B}^\dagger \hat{p}_1 \hat{B}=\frac{1}{\sqrt{2}}(\hat{p}_1+\hat{p}_2)\\
    \hat{B}^\dagger \hat{x}_2 \hat{B}&=\frac{1}{\sqrt{2}}(\hat{x}_1-\hat{x}_2),\hat{B}^\dagger \hat{p}_2 \hat{B}=\frac{1}{\sqrt{2}}(\hat{p}_1-\hat{p}_2).
    \end{split}
\end{equation}
Therefore, $v'$ can be expressed as
\begin{widetext}
\begin{equation}
    \begin{split}
        v'&=\frac{\int [(\xi-\tilde{\xi}(y))^2+(\eta-\tilde{\eta}(y))^2]\exp(-\frac{\xi^2+\eta^2}{v})W_{\hat{\sigma}}(\xi/\sqrt{2},\eta/\sqrt{2})d\xi d\eta}{\int \exp(-\frac{\xi^2+\eta^2}{v})W_{\hat{\sigma}}(\xi/\sqrt{2},\eta/\sqrt{2})d\xi d\eta}\\
        &=\frac{\int 2[(x-x_0)^2+(p-p_0)^2]\exp(-\frac{2}{v}(x^2+p^2))W_{\hat{\sigma}}(x,p)dxdp}{\int \exp(-\frac{2}{v}(x^2+p^2))W_{\hat{\sigma}}(x,p)dxdp},
    \end{split}
\end{equation}
\end{widetext}
where we put $x_0:=\tilde{\xi}(y)/\sqrt{2}, p_0:=\tilde{\eta}(y)/\sqrt{2}$.

When $v=2$, one can use the fact that
\begin{equation}
    \frac{1}{\pi}\exp(-(x^2+p^2))=W_{\ketbra{0}{0}}(x,p)
\end{equation}
and
\begin{equation}
\begin{split}
    &\frac{2}{\pi}[(x-x_0)^2+(p-p_0)^2]\exp(-(x^2+p^2))\\
    &=W_{\ketbra{0}{0}+\ketbra{\varphi}{\varphi}}(x,p),
\end{split}
\end{equation}
where $\ket{\varphi}=\alpha\ket{0}+\ket{1}$ and $\alpha=-\sqrt{2} (x_0+ip_0)$. Thus, we obtain
\begin{equation}
    \begin{split}
        v'&=\frac{\Tr[\qty(\ketbra{0}{0}+\ketbra{\varphi}{\varphi})\hat{\sigma}]}{\Tr[\ketbra{0}{0}\hat{\sigma}]}\\
        &\geq 1.
    \end{split}
\end{equation}

The equality holds when $\bra{\varphi}\hat\sigma\ket{\varphi}=0$ for some $\alpha$. The input and the POVM element being same and pure, \ie{} $\rhohat\propto\Ehat_y\propto \ketbra{\psi}{\psi}$ for some $\ket{\psi}$ is a sufficient condition for this, because
\begin{equation}
\begin{split}
\bra{1}\hat\sigma\ket{1}&\propto\int \qty|\int\phi_1(x_2)\psi\qty(\frac{x_1+x_2}{\sqrt{2}})\psi\qty(\frac{x_1-x_2}{\sqrt{2}})dx_2|^2 dx_1\\
&=0,
\end{split}
\end{equation}
where $\phi_1(x)$ is the wave function of $\ket{1}$. This holds because $\phi_1(x_2)$ is odd and $\psi\qty(\frac{x_1+x_2}{\sqrt{2}})\psi\qty(\frac{x_1-x_2}{\sqrt{2}})$ is even with respect to $x_2$.

\section{On the dependence of Eq.~(\ref{eq:vp_bound_final}) on post-selection probability $p(y)$}\label{section:smallp}
In Eq.~(\ref{eq:vp_bound_final}), the upper bound of $1/v'-1/v$ is inversely proportional to the normalized post-selection probability density $\frac{p(y)}{\sqrt{\Tr[\Ehat_y^2]}}$. We consider cases where $p(y)$ is small. Because $p(y)$ is the overlap between the prior distribution $p(\xi,\eta)$ (Eq.~(\ref{eq:gaussian_prior_dist})) and the likelihood function $p(\xi,\eta|y)$ (Eq.~\ref{eq:cond_prob}), this can be done by taking $y=y_0+ \epsilon$ for small $\epsilon$ where $y_0$ is a point such that $p(0,0|y_0)=0$ and taking $v\to 0$. We show that the linear dependence of the upper bound of $1/v'-1/v$ on the factor $\frac{\sqrt{\Tr[\Ehat_y^2]}}{p(y)}$ in Eq.~(\ref{eq:vp_bound_final}) is tight in the sense that we can construct an example such that
\begin{align}
    \lim_{\epsilon\to 0}\qty[\lim_{v\to 0}\frac{p(y)}{\sqrt{\Tr[\Ehat_y^2]}}]&= 0,\\
    \lim_{\epsilon\to 0}\qty[\lim_{v\to 0} (1/v'-1/v)\frac{p(y)}{\sqrt{\Tr[\Ehat_y^2]}}] &>0
\end{align}
for a small parameter $\epsilon$, with a fixed value of the maximum photon number $n$.

In the heterodyne setting in Sec.~\ref{section:non_gaussian}, we take $\rhohat=\rhohat'=\ketbra{1}{1}$. Then, for $y=(q,0)$, the estimation error $v'$ has the following expression:
\begin{widetext}
\begin{align}
    v'&=\frac{P_n(v,q)}{P_d(v,q)},\\
\begin{split}
     P_n(v,q)&=2 v (16 q^{8} + 32 q^{6} v^{2} - 128 q^{6} + 32 q^{4} v^{4} + 48 q^{4} v^{3}
     - 96 q^{4} v^{2} + 64 q^{4} v + 384 q^{4} + 16 q^{2} v^{6} + 80 q^{2} v^{5} + 160 q^{2} v^{4}\\
     & + 256 q^{2} v^{3}+ 256 q^{2} v^{2} - 256 q^{2} v
     - 512 q^{2} + 3 v^{8} + 20 v^{7} + 56 v^{6} + 112 v^{5} + 192 v^{4}+ 192 v^{3} + 128 v^{2} + 256 v + 256),
\end{split}\\
    P_d(v,q)&=\left(v + 2\right) \left(4 q^{4} + 4 q^{2} v^{2} - 16 q^{2} + v^{4} + 4 v^{3} + 8 v^{2} + 16 v + 16\right)^{2}.
\end{align}
\end{widetext}
On the other hand, the post-selection probability density $p(q,0)$ is expressed as

\begin{equation}
    p(q,0)=\frac{\left(4 q^{4} + 4 q^{2} v^{2} - 16 q^{2} + v^{4} + 4 v^{3} + 8 v^{2} + 16 v + 16\right) e^{- \frac{q^{2}}{v + 2}}}{\pi \left(v^{5} + 10 v^{4} + 40 v^{3} + 80 v^{2} + 80 v + 32\right)}
\end{equation}
If we take the limit of $v\to0$, we have
\begin{equation}
    \lim_{v\to0}\qty(\frac{1}{v'}-\frac{1}{v})=\frac{q^{4} - 4 q^{2} + 12}{2 \left(q^{2} - 2\right)^{2}}
\end{equation}
and
\begin{equation}
    \lim_{v\to0}p(q,0)=\frac{\left(q^{2} - 2\right)^{2} e^{- \frac{q^{2}}{2}}}{8 \pi}.
\end{equation}

Then, if we put $q^2=2+\epsilon$ and assume $|\epsilon|\ll 1$, we have $p(q,0)\sim \epsilon^{2}$ and $\qty(\frac{1}{v'}-\frac{1}{v}) \sim \epsilon^{-2}$. Using $\Tr\Ehat_y^2=1/(2\pi)^2$, we have
\begin{align}
    \lim_{\epsilon\to 0}\qty[\lim_{v\to 0}\frac{p(q,0)}{\sqrt{\Tr[\Ehat_{q,0}^2]}}]&=0,\\
    \lim_{\epsilon\to 0}\qty[\lim_{v\to 0} (1/v'-1/v)\frac{p(q,0)}{\sqrt{\Tr[\Ehat_{q,0}^2]}}]&=\frac{1}{e},
\end{align}
which is the desired property.
\clearpage
\bibliography{ref.bib}
\end{document}